\documentclass[letterpaper]{iopart}       

\usepackage{graphicx}

\begin{document}

\title{Symmetry breaking: A tool to unveil the topology
 of chaotic scattering with three degrees of freedom}

\author{C Jung$^1$, W P K Zapfe$^1$, O Merlo$^2$ and T H Seligman$^1$}

\address{$^1$ Instituto de Ciencias F\'isicas \\
Universidad Nacional Aut\'onoma de M\'exico, \\
Av. Universidad s/n, Apdo. Postal 48-3\\
$^2$Zurich University of Applied Sciences \\
  Institute of Applied Simulation \\
   Gr\"uental, P.O. Box 
           CH-8820 Waedenswil, Schweiz.\\ }
 
\ead{karelz@fis.unam.mx}

\bibliographystyle{unsrt}

\begin{abstract}

We shall use symmetry breaking as a tool to attack the problem 
of identifying the topology
of chaotic scatteruing with more then two degrees of freedom.
specifically we discuss the structure of the homoclinic/heteroclinic tangle and the
connection between the chaotic invariant set, the scattering 
functions and the singularities in the cross section for a class of scattering 
systems with one open and two closed degrees of freedom.

\textbf{ keywords:}
Chaotic Scattering, Cross sections, Classical Mechanics, Rainbow Singularities


\end{abstract}

\maketitle

\section{Introduction}
   
In a Hamiltonian system with two degrees of freedom the Poincare map acts on a 
two dimensional domain. The chaotic invariant set is represented by the well 
known horseshoe construction of Smale \cite{sma}. In the meantime it has been 
worked out quite well, how in the case of open systems (scattering systems)
the properties of this chaotic set are seen in scattering functions 
\cite{butik,jls} and in
scattering cross sections \cite{schelin,pue}.
 The corresponding investigations become a lot more
complicated when we proceed to systems with more degrees of freedom. Therefore
relatively little has been done for more degrees of freedom. In the present 
article we present some results for one class of higher dimensional scattering
systems, namely the one with one open and two closed degrees of freedom.

Our strategy to attack this higher dimensional case is based on an idea 
introduced to treat elliptical orbits in reduced three-body systems \cite{benet},
which is actually much in the spirit of Moshinskys work in nuclear physics.
We start from a system
with a symmetry and a conserved quantity (in the former case the 
Jacobi integral). Then the three degrees of freedom 
system can be reduced to a continuum of systems with two degrees of freedom 
where the numerical value of the conserved quantity serves as parameter. We 
use the knowledge  of the techniques for the two degree of freedom
case, to treat the reduced system. Then we embed the continuum of the reduced
systems into the full six-dimensional phase space to assemble the corresponding
higher dimensional structures still for the symmetric case. In other words, we find
the solution for the symmetric 
case.  Finally we break 
the symmetry to get to the true three degrees of freedom case. Here we 
show that the breaking of the symmetry does 
not have important effects
on the qualitative properties of the structure, at least as long as the 
symmetry breaking is weak. The argument is supported by 
numerical calculations.

\section{The chaotic invariant set}

For Hamiltonian systems with three degrees of freedom the Poincar\'e map acts on
a four dimensional domain. The first thing to do is to describe  in 
the domain of the four dimensional Poincar\'e map
the higher dimensional analog of the two dimensional horseshoe we obtain, 
when a symmetry reduces the effective dimensionality of the system.

As representative example of the class of system we want to treat we imagine
the motion of a particle in a channel with an obstacle. The empty channel is
cylindrical and thereby has a rotational symmetry in the azimuth angle $\phi$. 
The obstacle is represented by a short range potential
depending on a parameter A which describes the deviation of the obstacle from
rotational symmetry. For A=0 the obstacle is symmetric in azimuth direction 
and the conjugate angular momentum $L$
is conserved and the rotational symmetry allows the reduction to two degrees
of freedom. A natural choice for the intersection condition for the Poincare map
is the choice of relative maxima of the cylindrical radius $\rho$. Every times
when the particle runs through a relative maximum of $\rho$ we measure the
values of $z$, $p$, $\phi$ and $L$ where $z$ and $p$ are position and conjugate
momentum along the axis of the channel. The space of these four coordinates is 
the domain of the Poincar\'e map.

One possibility to proceed is to set up particular models for the channel and
the potential of the obstacle and to study solutions of Hamilton's equations
of motion and to obtain the Poincare map from them. We do something simpler
and more direct. We set up a model function for the Poincare map which serves
as prototype example (paradigm) for the whole class of systems. We compose 
it from three steps: \newline
First, half of a free flight \newline
Second, a kick generated by the generating function
\begin{equation}\label{genfun}
G(z,p',\phi,L')= z p' + \phi L' + (L_m - L')(1+A \cos(\phi)) V(z)
\end{equation}
Here $L_m$ is the maximally possible value of $L$, indirectly $L_m$ fixes
the value of the total energy $E$. \newline
Third, again half a free flight \newline
For the potential function $V$ appearing in Eq. \ref{genfun} we take 
\begin{equation}\label{potv}
V(z) = - \exp (-z^2)
\end{equation}

For $A=0$ the map becomes independent of $\phi$ and reduces to a
map acting on the $z$, $p$ plane. This reduced map depends on the conserved
value of $L$ as parameter.

Because the potential of  Eq. \ref{potv} is purely 
attractive the points $z= \pm \infty$
act as outer fixed points of the map and we use their invariant manifolds to
construct the horseshoe of the reduced system.

\begin{figure}
 \begin{tabular}{c}
  \includegraphics[width=0.95\textwidth]{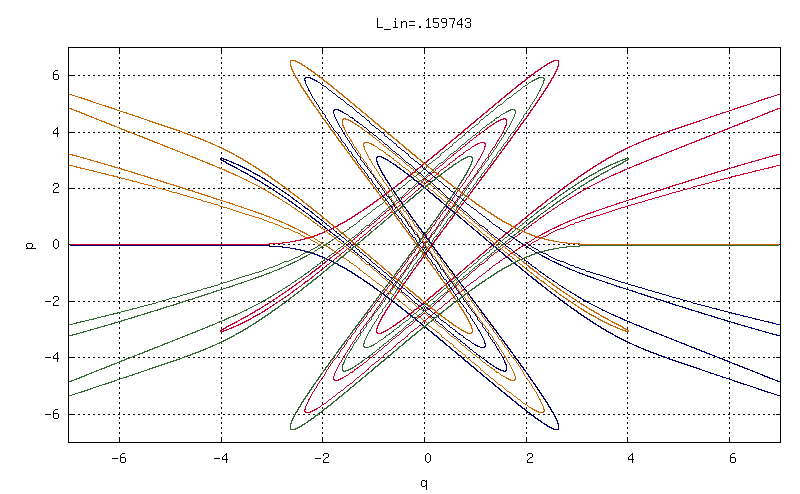} \\
  \includegraphics[width=0.95\textwidth]{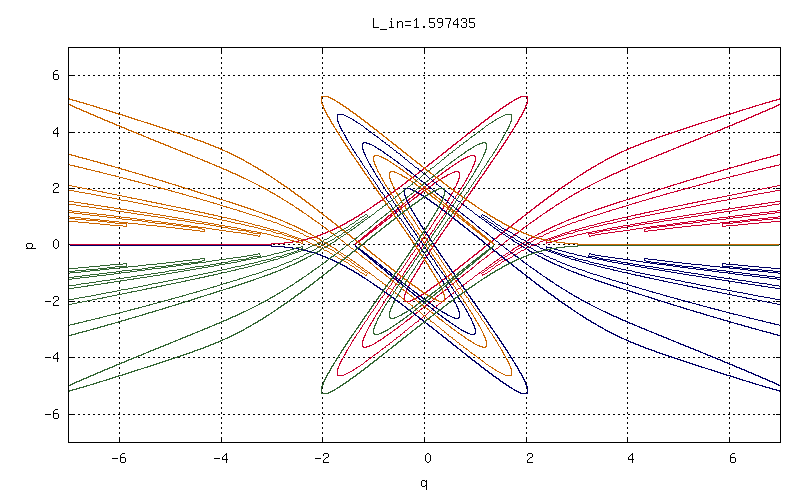} 
 \end{tabular}
 \caption{ In this figure and in the next two figures we show the horseshoe
of the reduced system for parameter value $A=0$. We show it for 6 different 
values of the conserved angular momentum. As can be seen, this angular
momentum acts as a development parameter for the horseshoe. For values of
$L_{in}$ close to zero ( as typical example the case $L_{in} = 0.159749 $ is 
seen in the upper part of the figure ) the horseshoe
is complete. For larger values of $L_{in}$ it becomes incomplete. The lower part
of Fig.1 shows a situation where the horseshoe has just become incomplete.}
 \label{develgama1}
\end{figure}

\begin{figure}
 \begin{tabular}{c}
  \includegraphics[width=0.95\textwidth]{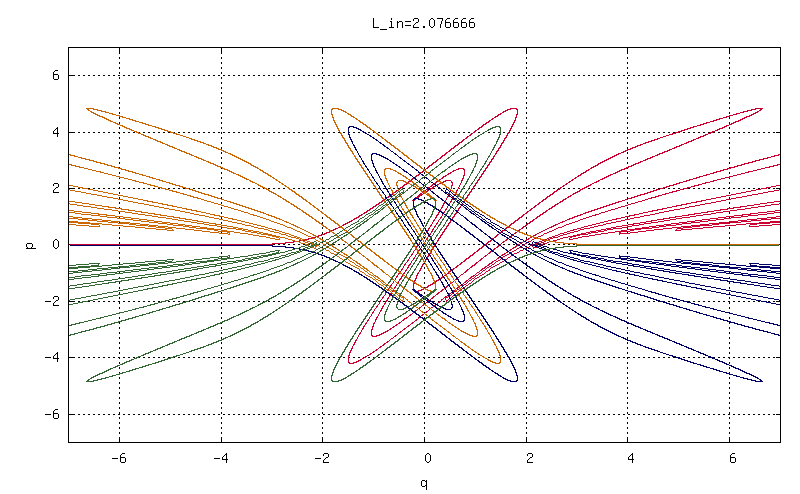} \\
  \includegraphics[width=0.95\textwidth]{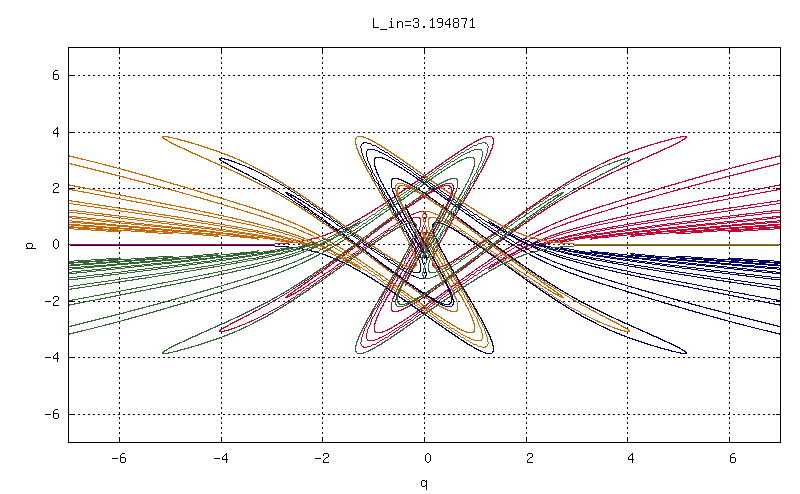} 
  \end{tabular}
 \caption{This figure is the continuation of Fig.1. The upper part shows the case
$L_{in}=2.076666$ where the horseshoe is a little more than 2/3 developed and
the lower part shows the case $L_{in}=3.194871$ where it is a little less than 
2/3 developed.}
 \label{develgama2}
\end{figure}

\begin{figure}
\begin{tabular}{c}
  \includegraphics[width=0.95\textwidth]{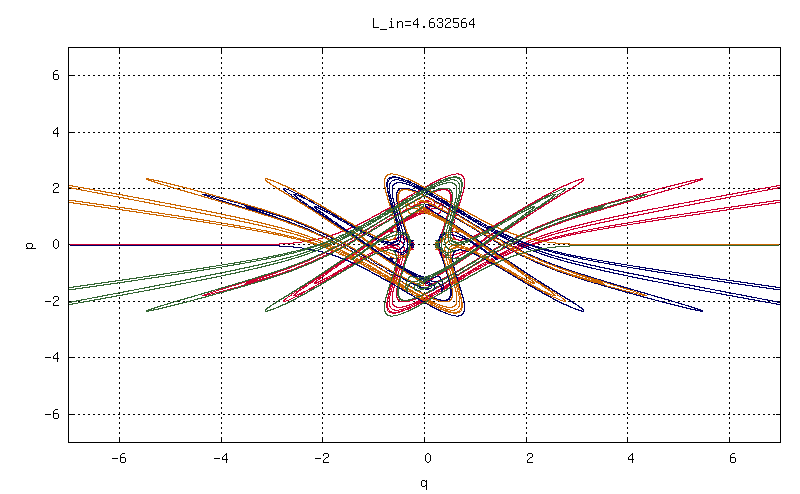} \\
  \includegraphics[width=0.95\textwidth]{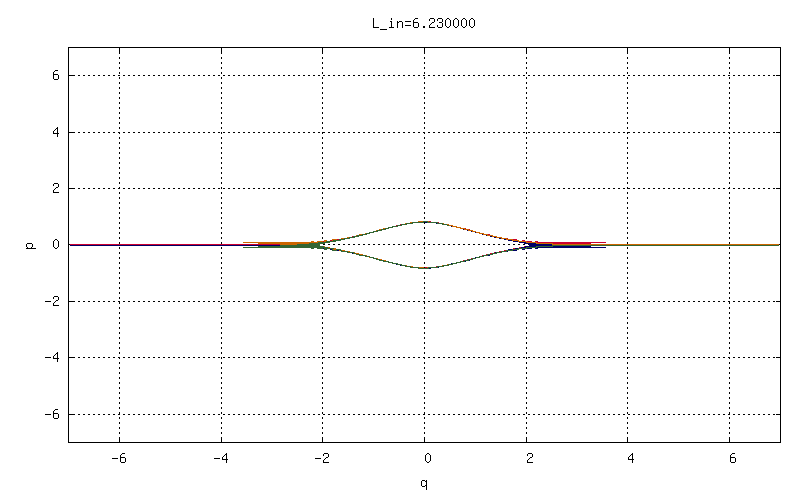}
\end{tabular}
\caption{This figure is the continuation of Figs.1 and 2. The upper part shows the
case $L_{in}=4.632564$ where the horseshoe is developed to the degree 
of almost one third.
The lower part shows a case where $L_{in}$ comes close to its maximally allowed
value $L_m$ at which the horseshoe development degree goes to zero, i.e. where
the horseshoe collapses to a parabolic line.}
 \label{develgama3}
\end{figure}

Figs. \ref{develgama1}, \ref{develgama2},  \ref{develgama3} shows the result for
various values of $L$. As function of $L$ we see the standard development
scenario of a ternary symmetric horseshoe. When $L$ goes from zero to 
approximately 5.5 the relative development stage of the homoclinic/heteroclinic
tangle shown in Figs. \ref{develgama1},  \ref{develgama2}, \ref{develgama3} 
goes from zero to one, for more details on the introduction
of a quantitative development parameter for a ternary symmetric horseshoe
see \cite{jls}.

The transition to the higher dimensional structures, still for the symmetric
case, is now easy. The angular momentum $L$ which serves as parameter in the
two dimensional map is a coordinate of the domain of the four dimensional map.
Therefore we just form the pile of the whole continuum of two dimensional cases.
This three dimensional structure is an intermediate step for the final four
dimensional structure. To include the final coordinate $\phi$ is completely
trivial. We just form the Cartesian product of the three dimensional intermediate
step with a circle representing the coordinate $\phi$. The result is the
final homoclinic/heteroclinic tangle in the four dimensional map. In this
process the outer fixed point at $z= \pm \infty$ of the two dimensional map
becomes a two dimensional continuum (two dimensional surface) of fixed points.
It is an example of what Wiggins calls a normally hyperbolic invariant manifold
(NHIM) \cite{wig}. This surface has one unstable direction, one stable direction
and two neutral directions. Its stable and unstable manifolds are three 
dimensional surfaces which serve as separatrix surfaces in the four dimensional
domain of the map.

Last we need an argument of structural stability. During the development scenario 
the two dimensional horseshoe runs through an infinity of tangencies. In the
higher dimensional homoclinic tangle we see instead transverse intersections, where
two surfaces of codimension one intersect in a curve of codimension two. This
intersection curve itself exists over a limited range of $L$ values only. At
$L$ values where the intersection curve disappears we see the tangency in the two
dimensional horseshoe. This transversality of the whole development scenario in
more dimensions leads to a corresponding structural
stability of the higher dimensional tangle. Under small deformations of the system,
and this includes breaking of the symmetry, we have qualitative changes in high
levels of the hierarchy only. At lower levels the tangle remains the same
qualitatively. This is the argument why the main structure of the higher
dimensional homoclinic tangle has a rather high robustness against breaking of the
symmetry until the deviation from symmetry becomes very large. 

\section{How the chaotic set appears in scattering functions}

The most useful scattering function for the present situation is obtained from
the following considerations. We label asymptotes by giving the total energy $E$,
the longitudinal momentum $p$, the angular momentum $L$ and two relative angles 
$\chi$ and $\psi$ between the radial and the longitudinal or the radial and the
azimuthal degree of freedom respectively. As scattering function we
give final momentum $p_{out}$ and the angular momentum transfer, the difference
between final and initial angular momentum $\Delta L = L_{out} - L_{in}$ as 
function of the two relative initial angles $\chi$ and $\psi$ for fixed values of
$E$, $L_{in}$ and $p_{in}$. In the case of $A=0$ this function is independent
of $\psi$, it always shows a fractal structure along $\chi$. 

\begin{figure}
 \begin{tabular}{c}
  \includegraphics[width=0.95\textwidth]{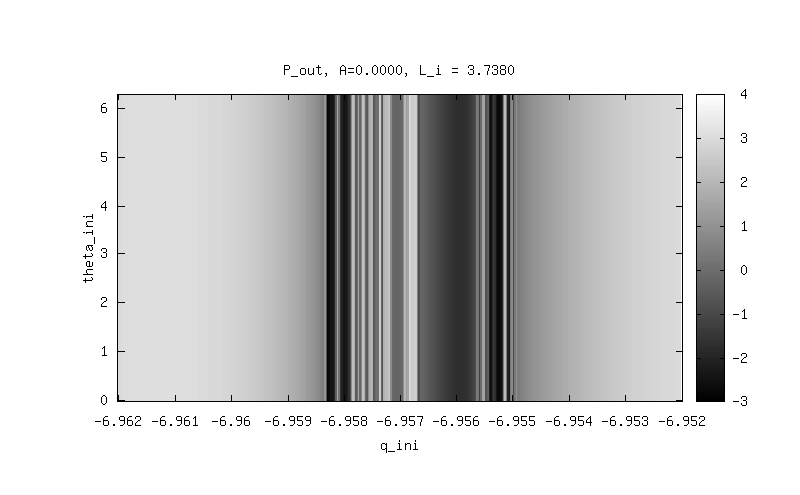} \\
  \includegraphics[width=0.95\textwidth]{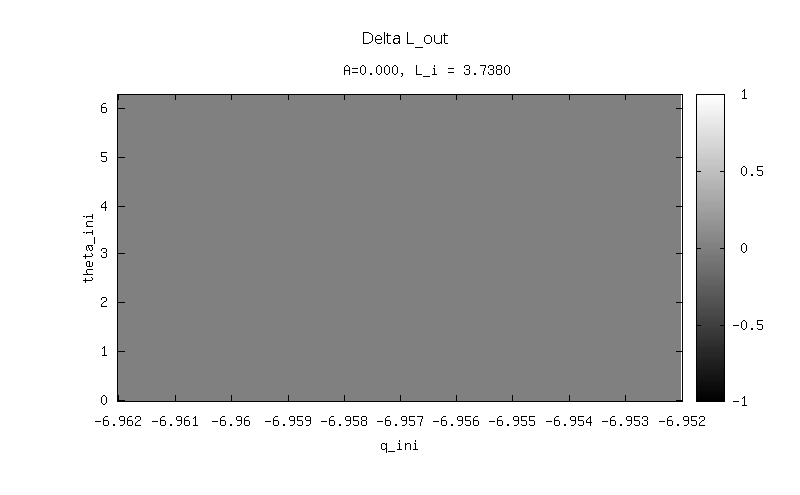} 
\end{tabular}
 \caption{This figure shows the scattering functions $p_{out}(\chi_{in},\psi_{in})$
in the upper part and $\Delta L (\chi_{in}, \psi_{in})$ in the lower part for 
the symmetric case $A=0$. In this symmetric case the angular momentum is a
conserved quantity and accordingly $\Delta L \equiv 0$, so the plot appears as 
homogeneous grey colour. The boundaries of the rectangular domain of these plots
should be identified to form a 2 dimensional torus. Note that the fractal
structure seen in the upper part of the figure is the Cartesian product of a
1 dimensional fractal in horizontal direction with a circle in vertical
direction.}
 \label{fundisp1}
\end{figure}

\begin{figure}
 \begin{tabular}{c}
  \includegraphics[width=0.95\textwidth]{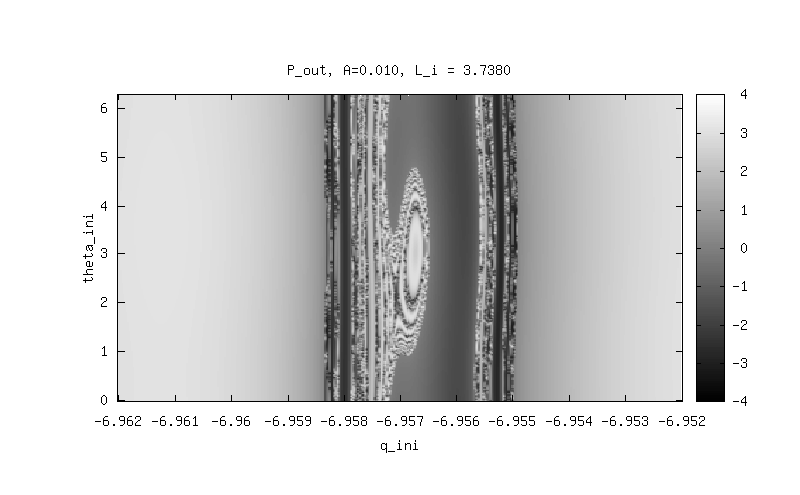} \\
  \includegraphics[width=0.95\textwidth]{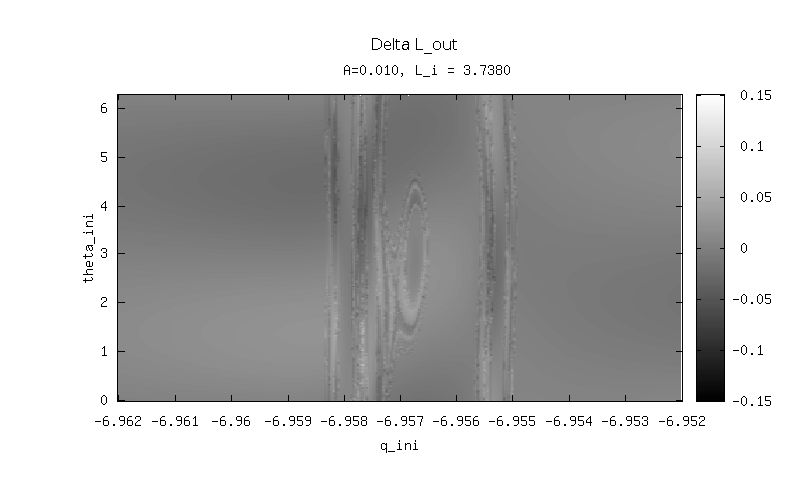} 
 \end{tabular}
 \caption{This figure shows the same scattering functions as Fig.4 for the small 
symmetry breaking $A=0.01$. Because of this symmetry breaking the fractal
structure has turned into a truly 2 dimensional fractal. However, a mayor
part of the intervals of continuity of the scattering functions still form
stripes running in vertical direction. These are qualitative features of the
product structure of the symmetric case which have survived the symmetry
breaking.}
 \label{fundisp2}
\end{figure}

\begin{figure}
 \begin{tabular}{c}
  \includegraphics[width=0.95\textwidth]{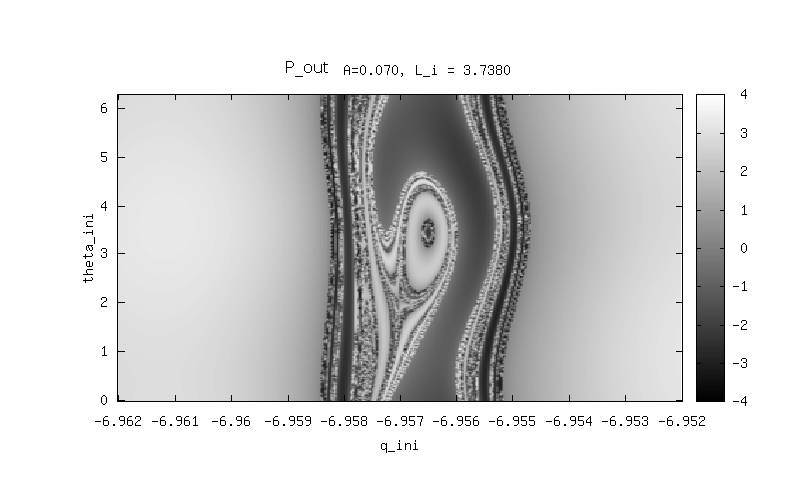} \\
  \includegraphics[width=0.95\textwidth]{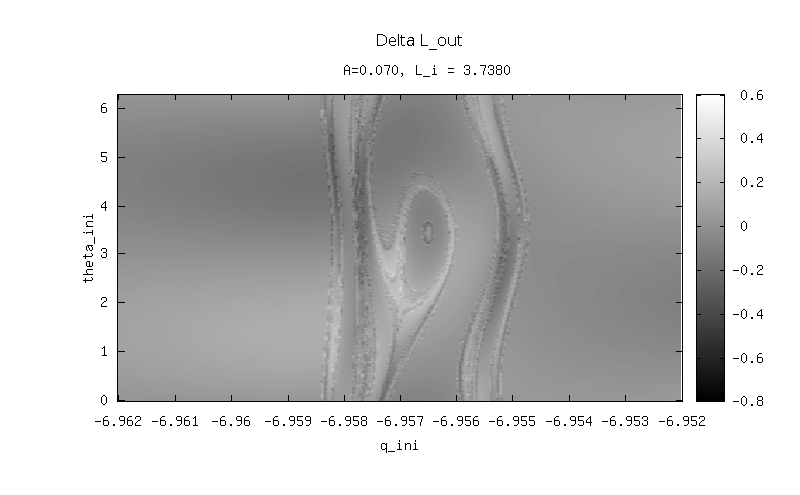}
 \end{tabular}
 \caption{This figure shows the same scattering functions as Figs.4 and 5 for
the moderate symmetry breaking $A=0.07$. The intervals of continuity show
stronger deformations compared to the symmetric case $A=0$ shown in Fig.4.
However, we still see many remnants of the stripes running in vertical
direction.}
 \label{fundisp3}
\end{figure}

In Figs. \ref{fundisp1}, \ref{fundisp2} and \ref{fundisp3} 
we show an example of this function for three values of A, namely
$A=0.0$ in parts a and b, $A=0.1$ in parts c and d and $A=0.3$ in parts e and f.
The scattering flow casts a kind of shadow image of the homoclinic tangle into
the outgoing asymptotic region and in the scattering functions we see exactly 
this shadow image. Singularities of the scattering function, i.e. boundaries
of intervals of continuity of this function indicate initial conditions on the
stable manifold of the NHIM at infinity. Thereby an analysis of the singularities 
of scattering functions allows the reconstruction of all the important information
concerning the chaotic set.  

Observe that for $A=0$ the scattering function has a factorization into a fractal
structure in $\chi$ direction and constancy in $\psi$ direction. Accordingly the
function shows intervals of continuity which are the product of an interval in
$\chi$ direction with a circle in $\psi$ direction. For $A$ small
only very tiny intervals of continuity of high level of hierarchy change their
qualitative structure. The larger intervals keep their basic topology of a strip
running around in $\psi$ direction. We have to go to the very strong symmetry
breaking of $A \approx 0.5$ to see the largest intervals change their qualitative
structure.

\section{How the chaotic set appears in the doubly differential cross section}

In this section we explain how the fractal chaotic set in phase space or in the
Poincar\'e map causes a fractal pattern of rainbow singularities in the cross section.
For the case of three degrees of freedom we generalize the ideas developed in
\cite{pue} for the case of two degrees of freedom.

Usually in scattering experiments we do not have any control over the relative 
phase shift variables $\chi$ and $\psi$. We keep $E$, $p_{in}$ and $L_{in}$
fixed and the angles $\chi$ and $\psi$ have a random distribution with constant 
density on a two dimensional torus. We measure the relative probability to find
values $p_{out}$ and $\Delta L$. This quantity properly normalized with
respect to the incoming flow is the doubly differential cross section. To 
calculate this cross section for a particular pair of values of $p_{out}$ and
$\Delta L$ we have first to find all the preimages $ (\chi_k, \psi_k)$ of the 
scattering function which lead to these particular values of the final variables.
Then we have to calculate the determinant
\begin{equation}\label{detk}
d_k =  det [ \partial (p_{out}, \Delta L) / \partial ( \chi_{in}, \psi_{in}) ]
\end{equation}
in the particular initial preimage point k. The cross section is given by
\begin{equation}
\frac {d^2 \sigma}  {dp d \Delta L} (p, \Delta L) = \sum_{preimages} 1 / | d_k |
\end{equation}

\begin{figure}
  \includegraphics[width=0.85\textwidth]{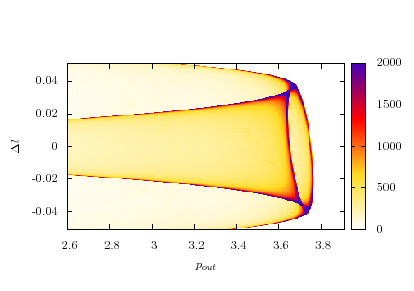}
  \caption{The contribution of a single interval of continuity of the scattering
functions ( it is the second largest one seen in Fig.5 ) to the doubly
differential cross section defined in Eqs. 3 and 4 for the case $A=0.01$.
Note that this figure coincides with the projection of half a torus or of a
ring shaped mountain. The caustics of this projection lead to one over square
root singularities ( rainbow singularities ) in the cross section.}
  \label{unasing}
\end{figure}

\begin{figure}
  \includegraphics[width=0.85\textwidth]{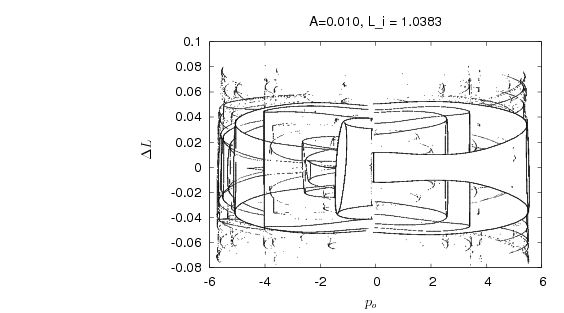}
  \caption{Singularity pattern seen in the doubly differential cross section, again
for the case $A=0.01$. In contrast to Fig.7 here the contributions from the
complete domain of the scattering functions are included. This pattern of
singularity lines is a fractal repetition of the basic pattern created by a
single interval of continuity as shown in Fig.7.}
  \label{varsing}
\end{figure}

We see immediately that the cross sections has singularities for such values of
$p_{out}$ and $\Delta L$ where the determinant of Eq. \ref{detk} is zero. These are values 
where several preimages collide and disappear, they are projection singularities
of the scattering function. These singularities are of one over square root type,
therefore the integral over the cross section is finite and there are no problems
with flux conservation. As long as the system is close to the symmetric case the
scattering function in each interval of continuity is similar to a ring shaped
mountain and accordingly the projection singularity of this function looks like
the one of half a torus (for details see \cite{jmsz}). Fig. \ref{unasing}
 shows the cross 
section contribution from a single interval of continuity. Fig.\ref{varsing} shows the
singularities from the whole toroidal domain of the scattering function. The
pattern seen in Fig.\ref{varsing}
 coincides with what a detector in a real experiment might
see. It is a fractal repetition of the basic singularity 
structure from Fig.\ref{unasing}.

We can change a parameter of the system, this might be the symmetry breaking 
parameter, or a action like initial condition, this might be the incoming angular
momentum $L_{ini}$. When we measure the cross section singularities as function of
this parameter then we can follow in the cross section the changes of the chaotic 
set under these changes of the system. This is an interesting contribution to the
inverse chaotic scattering problem.

\section{Conclusions}  

The stable manifold of a NHIM at infinity consists of trajectories which go out
to infinity while at the same time the longitudinal momentum goes to zero, they are
trajectories which arrive at infinity with velocity zero. These manifolds are
separatrix surfaces between trajectories going out to infinity and trajectories
which return to the interaction region at least once more before eventually also
going out to infinity. The trajectories on the unstable manifolds show the same
behaviour under time reversal. The homoclinic/heteroclinic tangle between the
stable and unstable manifolds of the NHIM at $\pm \infty$ are the higher dimensional
generalizations of the horseshoes in two dimensional maps. In our class of systems
this chaotic structure exists for all values of the total energy $E$. The reason is
that in our class of systems for any value of $E$ there are trajectories going out
to infinity with longitudinal velocity zero. This happens thanks to the closed
degrees of freedom which can swallow up any amount of energy such that only energy
zero is left for the open degree of freedom. In this sense our system is at a channel
threshold of the inelastic scattering for any value of $E$.

Because of these considerations it is clear that we do not expect a
similar behaviour for scattering systems with open degrees of freedom only. For
a discussion of this point for scattering off four hard spheres see \cite{chen}.
The Poincare map of this system has hyperbolic fixed points but does not have
any NHIM. 

For systems with more than three degrees of freedom similar considerations hold.
For each additional degree of freedom the NHIM and its manifolds acquire two
additional neutral directions. The domains of the scattering functions and of the
cross section acquire one additional dimension.

\section*{Acknowledgement}

This work has been supported by CONACyT under grant number 79988 and by
DGAPA under grant number IN-110110

\end{document}